# Instantly Obsoleting the Address-code Associations: A New Principle for Defending Advanced Code Reuse Attack

Ping Chen, Jun Xu, Jun Wang, Peng Liu

*College of Information Sciences and Technology,Pennsylvania State University*

`{pzc10, jxx13, pliu}@ist.psu.edu`

Fine-grained Address Space Randomization has been considered as an effective protection against code reuse attacks such as ROP/JOP. However, it only employs a one-time randomization, and such a limitation has been exploited by recent just-in-time ROP and side channel ROP, which collect gadgets on-the-fly and dynamically compile them for malicious purposes. To defeat these advanced code reuse attacks, we propose a new defense principle: *instantly obsoleting the address-code associations*. We have initialized this principle with a novel technique called *virtual space page table remapping* and implemented the technique in a system CHAMELEON. CHAMELEON periodically re-randomizes the locations of code pages on-the-fly. A set of techniques are proposed to achieve our goal, including iterative instrumentation that instruments a to-be-protected binary program to generate a re-randomization compatible binary, runtime virtual page shuffling, and function reordering and instruction rearranging optimizations. We have tested CHAMELEON with over a hundred binary programs. Our experiments show that CHAMELEON can defeat all of our tested exploits by both preventing the exploit from gathering sufficient gadgets, and blocking the gadgets execution. Regarding the interval of our re-randomization, it is a parameter and can be set as short as 100ms, 10ms or 1ms. The experiment results show that CHAMELEON introduces on average 11.1%, 12.1% and 12.9% performance overhead for these parameters, respectively.

# 1 Introduction

Code reuse attacks, such as return-oriented programming (ROP)[40, 12, 9, 37], recently have been used to exploit vulnerabilities in widely used commercial software such as Internet Explorer [26] and Adobe Reader [17]. ROP attacks hijack program control flow to execute code snippets (i.e., "gadgets") that are elaborately selected from shared libraries or the application code. Without injecting any code, ROP can easily bypass data execution prevention (DEP) [30, 31]. There have been two major lines of research towards defeating ROP: Control Flow Integrity (CFI) [1, 23, 44, 47, 18, 48, 8] and Address Space Randomization (ASR) [4, 5, 46, 36, 43, 45, 24, 34, 27, 3].

Recent efforts on ASR have been pushing the randomization to be more fine-grained [45, 24, 34, 27, 3] to defend advanced ROP attacks (e.g., [37]). Unfortunately, there is an emerging trend in the attacker community: ROP attacks are launched together with memory disclosure attacks [13, 42, 39]. This results in a new type of "one-two punch" ROP attacks. A real world "one-two punch" ROP of this type is the just-in-time ROP (JIT-ROP) [41]. Besides memory disclosure attacks, side channel attacks can also be combined with traditional ROP attacks to form another type of "one-two punch" ROP attacks (we call SC-ROP) [7, 38]. "One-two punch" ROP attacks can circumvent the most advanced ASR techniques, including fine-grained ASR. Before explaining why the "one-two punch" ROP attacks are hard to be defended, we would like to give a short introduction of these attacks.

A JIT-ROP attack first exploits a memory disclosure vulnerability to get one page of memory. By disassembling the initial page, the attacker aims to find local gadgets. If local gadgets are insufficient, the attack will (1) find locations of other pages by analyzing the current page; (2) exploit the memory disclosure vulnerability again to get those pages; (3) scan the newly obtained pages to identify more gadgets. Note that the attack does not continue scanning pages next to the current page to avoid causing problems such as segmentation fault. Finally, the attack "compiles" the start addresses of the identified gadgets and the data these gadgets will use into a malicious payload.

SC-ROP attacks collect gadgets *on-the-fly* in a different way from JIT-ROP attacks. By exploiting a vulnerability such as buffer overflow, a SC-ROP attack can redirect the execution to a specific location. The execution behaviors after the redirection enable the attack to infer what instructions have been executed. Useful behaviors include execution status (e.g., crash or hang-up) [7], execution results (e.g., program output) [38], and execution time [38]. By iteratively changing the redirecting locations and inferring the executed instructions, the attack can understand *where* valid instructions are located and *what* these instructions are.

We can notice that the above two types of "one-two punch" ROP attacks share a common feature that they identify *address-code associations* on-the-fly. For a process, we say an address $A$ and an instruction $I$ forms an address-code association if $I$ resides at logical address $A$. Existing ASR techniques randomize the address-code associations for one time before the execution (e.g., at load-time). Such a strategy randomizes the mapping from the code layout in the binaries on disk to code layout in the memory and thus, raises the bar for traditional ROP attacks that rely on the mapping information. However, "one-two punch" ROP attacks identify the address-code associations during execution time and doing this requires no knowledge about the mapping. This leads to the fact that "one-two punch" ROP attacks bypass existing ASR or fine-grained ASR techniques. Making the randomization granularity finer and finer is not an effective principle to change the fact.

For ASR techniques to defeat "one-two punch" ROP attacks, a fundamentally new principle is required. In this paper, we propose such a new defense principle, namely, *Let the address-code associations quickly become obsolete*. There are two implications from this principle:

- **Implication 1**: the address-code associations used by "one-two punch" ROP attacks should become obsolete before the ROP shellcode is executed.

- **Implication 2**: the legitimate binary should be able to use the valid address-code associations, no matter how frequently the associations keep changing.

The new principle is inspired by two weaknesses of an "one-two punch" ROP attack. First, the attack needs an unavoidable time window to achieve its goal. The time window consists of two parts: ($t_1$) the time the attack takes to identify the address-code associations; ($t_2$) the time the attack takes to construct the malicious payload. Second, the address-code associations used to construct the payload must be valid until the execution of ROP payload (i.e., the execution is directed by the payload). Otherwise, the instructions residing in the addresses would not be the intended ones when the ROP payload is executed.

The two weaknesses imply two properties of an ASR technique that enforces the new principle. First, how quickly the ASR technique should obsolete the address-code associations is determined by the unavoidable time window (i.e., $t_1 + t_2$). Second, the "one-two punch" ROP attack is guaranteed to fail when the ASR is deployed. The reason is that the first implication of the principle takes advantage of the second weakness of "one-two punch" ROP attacks. However, to our best knowledge, there exists no ASR techniques enforcing the new principle.

In this paper, we propose a novel ASR technique, *virtual address page table remapping* (VAPTR), to enforce the new principle. VAPTR periodically randomizes the locations of code pages *on-the-fly*. Essentially, the address-code associations become *obsolete* after each round of re-randomization. Our experiments show that VAPTR can make the obsoleting cycle quicker than the unavoidable time window of "one-two punch" ROP attacks. We have developed a system named as CHAMELEON[1] to implement the VAPTR technique. Overview of the CHAMELEON system is presented in §2.3.

The main contributions of our work are as follows.

- We propose a new fundamental defense principle for ASR techniques to defeat "one-two punch" ROP attacks. Enforcement of the principle in ASR techniques is guaranteed to be invincible by 'one-two punch" ROP attacks.

- We propose a novel VAPTR technique to enforce our defense principle. This technique involves a new type of binary instrumentation.

- We have developed the CHAMELEON system to implement the VAPTR technique. Experimental results with over a hundred binary programs show that CHAMELEON is able to defeat "one-two punch" ROP attacks with very low overhead, for instance: httpd with 9.1%-11.7%, lighttpd with 13.8%-16.7%, and nginx with 18.1%-20.7% when setting the randomization intervals to be 100*ms* and 1*ms*, respectively.

## 2 System Overview

### 2.1 Threat Model

In the threat model, we assume a program runs in a Linux operating system with DEP and ASLR enabled. The victim program contains a memory corruption vulnerability. The attacker aims at exploiting this vulnerability to execute arbitrary malicious logic through ROP. The attacker is assumed to be able to: (1) access the executable (or the source codes) of the program; (2) control major input channels to the process, including standard inputs, file streams, and network traffics; (3) identify address-code associations at runtime, either via memory disclosure attacks or side channel attacks; (4) find sufficient useful instructions (or gadgets). However, the attacker has not gained the root privilege to access the OS kernel. In addition, the attacker needs an unavoidable time window to identify address-code associations and construct the payload.

---

[1]CHAMELEON stands for CHAnge the MEmory Layout pEriodically and ON-the-fly.

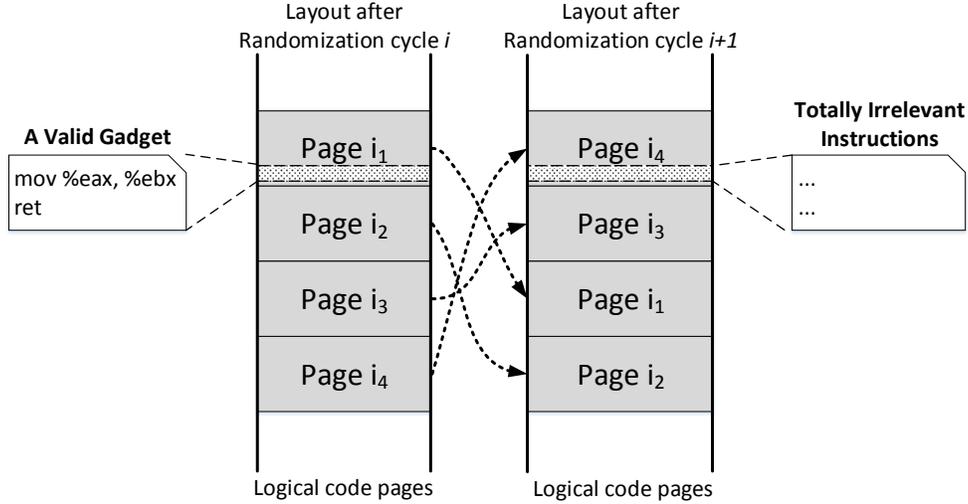

Figure 1: High level idea of VAPTR.

## 2.2 Approach Overview

At a high level, our VAPTR technique periodically randomizes the *mapping* between the logical page numbers and the physical page numbers of code pages. We keep the physical address space unchanged and shuffle the code pages in the logical address space during each cycle of randomization. As presented in Figure 1, part of the layout of the logical address space after cycle $i$ and cycle $i+1$ are shown as the left part and right part, respectively. Note that the code inside a physical page (e.g., Page $i_1$) will never be moved to another physical page. So the dotted arrow between the left part and the right part means the codes in a page (e.g., Page $i_1$) are kept unchanged, but the logical addresses of the codes are changed (e.g., logical addresses of codes in Page $i_1$ are changed from being inside the first logical page to being inside the third logical page).

When the randomizing cycle is shorter than the unavoidable time window, VAPTR follows the new defense principle and is guaranteed to defeat the threat model. To implement the VAPTR technique and enforce the new principle, several problems must be resolved. We briefly present these problems and our solutions as follows.

**Where and how to generate the randomized mapping information.** Technically, the randomized mapping information can be generated either in the user space or the kernel space of a process. However, generating such information in the user space leads to two attack surfaces. First, the randomized mapping information is maintained in the user space. The attacker can launch memory disclosure attacks to get such information and bypass the defense. Second, the mapping information generator code reside in the user space. The attacker might reuse the generator to learn the randomization logics or even perform de-randomization. So we choose to generate the randomized mapping information in the kernel space.

Regarding how to generate the randomized mapping information, there are two main options. First, we can modify the page table but find a way to avoid crashing the program. Second, we do not modify the page table but find a way to periodically modify the logical addresses used by the program. We choose the first option and propose a novel way to avoid crashing the program.

During each randomization cycle, we modify the page table to randomly generate a new bijection between the logical page numbers and the physical page numbers. As for how to avoid crashing the program, we present the solution in the next paragraph.

**How to maintain the normal execution of the program.** If we do not instrument the program code,

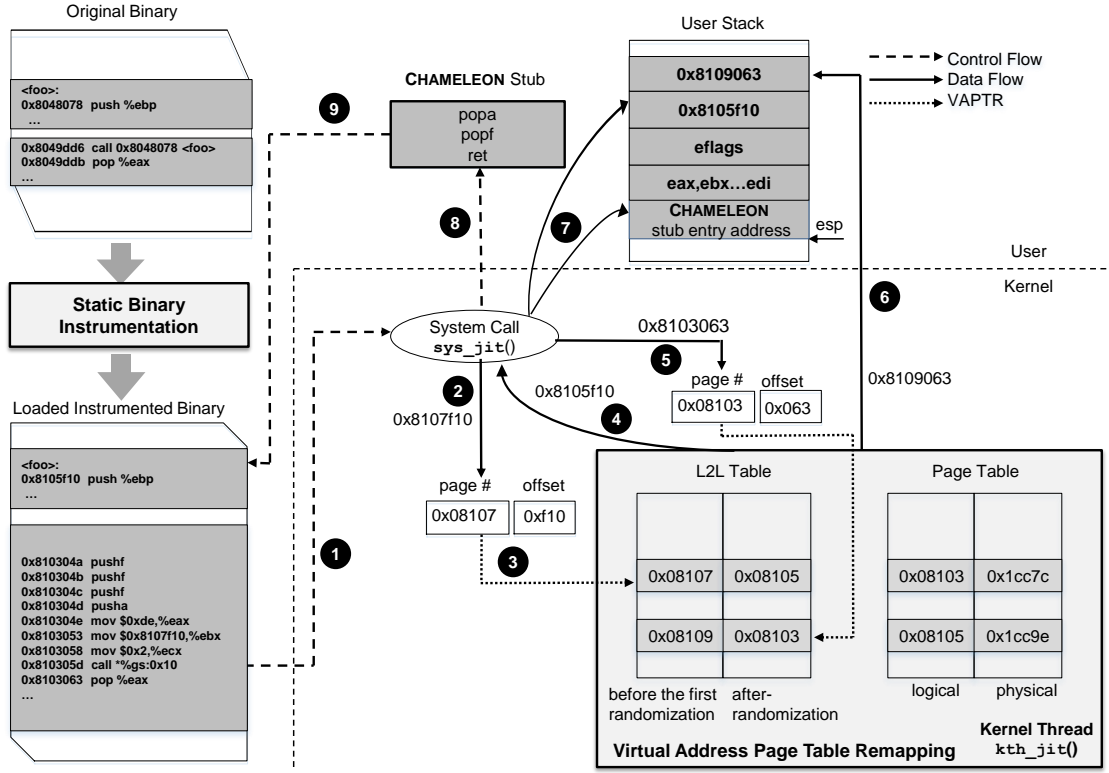

Figure 2: An Overview of CHAMELEON.

whenever the page table is modified, the program will crash[2]. So the primary issue is how to instrument the code so that arbitrary page table modification is accommodated. We address this issue in a novel way. In particular, we add one layer of indirection between the program code and the page table.

This layer of indirection is implemented through a second table in the kernel space, which is the *logical to logical mapping table* (L2L table). For a code page, its logical page number before the first cycle of the randomization is defined as the before-randomization logical page number. After that, the logical page number of the page is called as after-randomization logical page number. Note that the after-randomization logical page number changes with time. The L2L table records the mapping between the before-randomization logical page numbers and the after-randomization logical page numbers. The L2L table and the page table will form a two-step mapping. A before-randomization logical page number can be mapped to the correct physical page number through the two-step mapping. The L2L table plays a critical role in our solution. On one hand, it enables the page table to be modified in arbitrary way. On the other hand, it enables all programs to be instrumented in an unified way.

The remaining issue is how to instrument the program to use the L2L table. Our approach is to statically rewrite the binary executables. The rewriting is specific to the types of instructions. We classify instructions into two types: *inter-page instructions* and *intra-page instructions*. Inter-page instructions refer to control flow transfer instructions whose target addresses go across the current code page. In this paper, we regard control flow transfer instructions as the set of (a) direct `jump`, (b) indirect `jump`, (c) conditional `jump`, (d) direct `call`, (e) indirect `call`, and (f) `ret`. Instructions other than inter-page instructions are classified as intra-page instructions.

---

[2]If we do not instrument the program code, we need to make the kernel intercept each instruction. A possible way is to enforce the execution of any instruction to cause a page fault exception.

During the rewriting, we replace each inter-page instruction with a customized system call[3]. In our design of VAPTR, we enforce the target address of any inter-page instruction to be the before-randomization value (see §3.2.5 for details). The system call will use L2L table to translate the page number of the before-randomization target address. For intra-page instructions, we keep them unchanged and guarantee that the execution will never go wrong (please refer to §3.1.1).

**How to defeat the attacker.** Suppose the attacker selects to launch a JIT-ROP attack. During the attack, the attacker gets a gadget at address $A_1$ via memory disclosure attacks. The address $A_1$ later on becomes part of the ROP payload. In total, the attacker spends $T_w$ on getting the gadgets and constructing the payload. When the payload is getting executed, $A_1$ will be translated through the page table. However, during $T_w$, the page table is randomized at least once. To this end, $A_1$ will be translated to an irrelevant physical address during the execution of the ROP payload. The situation of defending against SC-ROP attacks is similar.

The attacker might extend the JIT-ROP attacks through reusing our system call. How we handle such extended attacks will be discussed in §3.2.6.

## 2.3 Chameleon Architecture Overview

We develop CHAMELEON to implement the VAPTR technique. CHAMELEON consists of two components: a static binary rewriter and a kernel space VAPTR unit.

Given a to-be-protected binary program, the rewriter instruments the binary to make it compatible with VAPTR. Specifically, the rewriter will replace each inter-page instruction with our system call `sys_jit`. At runtime, the instrumented program will invoke the system call to use the L2L table for address translation.

The kernel space VAPTR unit consists of three components: a kernel thread `kth_jit`, the L2L table, and a customized system call `sys_jit`. `kth_jit` periodically randomizes the page table and updates the L2L table correspondingly. When the instrumented program needs to use the L2L table for address translation, it will invoke the `sys_jit`. `sys_jit` queries the L2L table for translating the address and then transfers the execution back to the user space.

**A Working Example.** To demonstrate how CHAMELEON exactly works, we present a working example in Figure 2.

At first, the static binary rewriter generates a re-randomization suitable binary (RSB). In this example, the original binary invokes a `call 0x08048078` at `0x08049dd6`, which is an inter-page instruction. This instruction calls the function `foo`. The static binary rewriter replaces this `call` with 8 instructions. One of these instructions calls `sys_jit` at `0x0810305d`. Details of other instructions are explained in §3.1.2.

During the execution of the RSB, `sys_jit` will be invoked when the `foo` is to be called. The before-randomization logical target address `0x08107f10` is passed as an argument to `sys_jit` (❶). Then `sys_jit` splits `0x08107f10` into two parts (❷): the page number `0x08107` and the offset `0xf10`.

`sys_jit` looks the page number up in the L2L table to get the after-randomization page number `0x08105` (❸). Adding the offset `0xf10` to `0x08105`, `sys_jit` can get the after-randomization address `0x08105f10` (❹).

In addition, `sys_jit` will translate the return address to the before-randomization logical address, and write it to a reserved space in the current stack frame (❺ and ❻). Reason of such a recovery is explained in §3.2.5. Finally, `sys_jit` has to transfer the execution to the after-randomization address `0x8105f10` (❼, ❽ and ❾). Details can be referred to §3.2.2.

---

[3] There could be alternatives of system call as long as we can trap to the kernel, e.g., `int3`.

# 3 Design and Implementation

In this section, we present the design and implementation of the CHAMELEON system. Following the working process of the system, we first present how we design the static binary rewriter and then explain how we design the VAPTR unit to work with an instrumented binary.

## 3.1 Static Binary Rewriter

The goal of the rewriter is to instrument a normal binary program to be compatible with VAPTR. In this section, we describe how we design the rewriter to achieve such compatibility. First, we explain details of the instrumentation process of using our rewriter (§3.1.1). Second, we summarize the challenges we faced and how we address them in the instrumentation process (§3.1.2, §3.1.3). Third, we present the optimizations to improve the performance of the instrumented binary (§3.1.4).

### 3.1.1 Instrumentation Process

For different type of instructions, our rewriter takes different instrumenting actions.

In our design, CHAMELEON only randomizes the mapping information for code pages. In particular, CHAMELEON does not re-randomize the current code page which contains the instructions being executed, and will continue re-randomizing this page when the execution goes to another page. As such, we do not instrument intra-page instructions. Here we explain that not instrumenting intra-page instructions does not make the execution go wrong. Suppose $I_1$ and $I_2$ are two instructions executed sequentially in code page $P$ and $I_1$ is an intra-page instruction. Under VAPTR, the logical addresses of codes in $P$ are mapped to the fixed physical addresses. The distance between the physical addresses of $I_1$ and $I_2$ equals to the distance between their logical addresses. This naturally maintains the normal execution.

For each inter-page instruction, the rewriter replaces it with a *RSI unit*. A RSI unit is a set of customized instructions which mainly prepare two arguments for sys_jit and invoke sys_jit. More details are presented in §3.1.2. In addition, the rewriter inserts a RSI unit at the end of each code page and moves the instructions whose positions are occupied by the unit to the beginning of the next page.

We build the rewriter on the top of Dyninst[25]. Dyninst can help us identify the inter-page instructions. However, we face several challenges as follows.

**Register Conflicts.** Registers used by the neighboring instructions in the instrumentation point might also be used in a RSI unit. We explain how we handle such register conflicts in §3.1.2.

**Memory Expansion.** The inserted RSI units will change the code layout and result in memory expansion. By memory expansion, we mean two situations. First, an intra-page instruction is "pushed" into the next page. If this instruction is a control flow transfer instruction, it becomes an inter-page instruction. Second, on the other way around, an inter-page instruction can be changed into an intra-page instruction due to such "pushing". We explain how we handle such memory expansion in §3.1.3.

**Position Independent Code.** sys_jit needs to work in a special way when it is invoked in Position Independent Code (PIC). When instrumenting PIC, the rewriter sets the flag argument of sys_jit to a special value to indicate the sys_jit is called in PIC and in which dynamic library. VAPTR will use a special approach to handle PIC code (§3.2.4).

**Callback functions.** Callback functions in an instrumented program might be used by un-instrumented code, such as the OS kernel. However, the un-instrumented code will not "know" the after-randomization addresses of callback functions. To handle this issue, we first identify signal handlers and callback functions that are explicitly registered in other functions such as `atexit`. Then we separate the entry and body of each callback function. Entries of all callback functions are moved to the same page and we never randomize this page; while pages containing the bodies are randomized. We place a RSI unit at the entry of each callback function. When a callback function is invoked, the corresponding RSI unit will be executed and redirect the control flow to the body of the callback function.

### 3.1.2 RSI Unit Construction

A RSI unit needs to accomplish three goals: (1) reserving the memory space, (2) saving registers' contents, and (3) invoking our `sys_jit` system call.

**Reserving Memory Space.** A RSI unit needs to reserve some spaces on the user stack. These spaces will be used by `sys_jit`. We use `pushf` for space reservation. If we use `sub $0x8, %esp`, the flag register will be affected.

For different inter-page instructions (replaced by the RSI unit), the RSI unit needs to enforce different policies regarding the space reservation. Specifically, for `call`, we reserve two words; for `ret`, it does not need any reservation; for `jmp/jcc` and RSI unit at the end of a code page, we reserve one word. How `sys_jit` uses the reserved space will be illustrated in §3.2.2.

**Saving Registers Content.** The RSI unit needs to save the flag register and all the general purpose registers into the stack. The purpose is two-fold. First, we need to recover the registers upon exit of `sys_jit` to avoid register conflicts caused by the RSI unit. Second, `sys_jit` might need the values of registers during address translation (e.g., the saved `eflgas` can help determine the target address of a replaced `jcc`). We use `pushf` to save the `eflags` register and `pusha` to save all the general purpose registers.

**Invoking `sys_jit`.** `sys_jit` takes two arguments: the before-randomization target address of the replaced inter-page instruction and an additional flag. The flag indicates the type of the replaced instruction, such as `call` instruction. The RSI unit uses three `move` instructions to set the system call number and the two arguments. In particular, we use `call %gs:0x10` to invoke `sys_jit`.

**Example.** In Figure 2, the `call` instruction is replaced by 8 instructions: reserve the stack space (the first two `pushf`); store register state (the third `pushf` and `pusha`); set up the system call number (`mov $0xde, %eax`); set up the two arguments (`mov $0x8107f10, %ebx; mov $0x2, %ecx`); invoke the system call (`call *%gs:0x10`).

### 3.1.3 Iterative Instrumentation

To address the memory expansion problems, we propose two approaches: (1) we can first insert one RSI unit before each control flow transfer instruction, including `jcc, jmp, call, ret`. Then we identify all intra-page instructions replaced by RSI units. We recover these RSI units back to the original control flow transfer instructions and pad the redundant spaces with `NOP` instructions; (2) we iteratively instrument inter-page instructions in the binary program until we reach a fix point. Each round of instrumentation is conduced over the original clean binary but analysis of inter-page instructions is based on the layout of instrumented binary in the previous iteration. The fix point being reached means that the instrumented inter-pages instructions in the current iteration are identical to those instrumented in the previous iteration.

The first method is simple and quick. However, such an instrumentation strategy unnecessarily adds `NOP` for intra-page instructions. More importantly, this strategy does not support rearranging the binary layout, which prohibit us from performing extra optimizations, such as the two presented in §3.1.4.

We choose to use the second approach and denote it as "iterative instrumentation". Iterative instrumentation saves the unnecessary space for instrumenting intra-page instructions and support our optimizations.

A side benefit of our instrumentation is that all intended inter-page instructions are replaced with our RSI units. This reduces the number of available `call, ret, and jmp` for ROP attacks and increases the difficulty for ROP attackers.

### 3.1.4 Performance Optimization

Each instrumented inter-page instruction needs to trap into the kernel space, which introduces significant costs due to context switch. Reducing the number of inter-page instructions is therefore a good strategy to optimize the performance.

Two key observations enable us to do such optimization: (1) many functions have less than one page of code but spans two pages in the address space. These functions often contain inter-page instructions whose target addresses are still in the function. We denote these inter-page instructions as type-I optimizable inter-page instruction; (2) in functions with more than one page of code, many inter-page instruction's target addresses are within 4096 bytes and still in the functions. We denote these inter-page instructions as type-II optimizable inter-page instructions.

We propose two optimizations to remove the above two types of inter-pages instructions: function reordering and instruction rearranging. Function reordering works on functions with less than 4096 bytes of code (type-I function); instruction rearranging works on functions with more than 4096 bytes of code (type-II function).

**Function reordering.** The purpose of function reordering is to move functions with less than one page of code into a single memory page. Thus, type-I optimizable inter-page instructions are eliminated. The optimization is completed in five steps. First, we instrument all inter-page instructions in the first iteration of instrumentation. Second, in each follow-up iteration, we identify and rank type-I functions based on their code size. Third, we move all type-I functions into lower memory space than type-II functions. Fourth, we rearrange type-I functions into code pages starting from the code segment and prevent any type-I function from occupying two pages. Finally, for each page, we put one RSI unit at the end of the page to make sequential execution through two pages correct.

**Instruction rearranging.** For type-II optimizable inter-page instructions in type-II functions, we rearrange these instructions via inserting padding bytes (e.g., `NOP`) before them. We control the padding to make the target address and the address of a type-II optimizable inter-page instruction in the same page. To this end, type-II optimizable inter-page instructions are removed.

## 3.2 Virtual Address Page Table Remapping

The goal of our VAPTR component is to provide two services: a mechanism to perform the randomization periodically, and a way to let user space process to query the L2L table and then redirect the control flow to the target address. Our `kth_jit` kernel thread (§3.2.1) is designed to achieve the first goal and the `sys_jit` system call (§3.2.2) together with our CHAMELEON stub are designed for the second goal. Since both `sys_jit` and `kth_jit` are executed inside the kernel, we also have to take care of their synchronization (§3.2.3). In §3.2.4 and §3.2.5, we explain how we handle Position Independent Code and dynamically computed addresses, which are two challenges faced by us. Finally, we present how to secure `eip` on stack and disable the use of `sys_jit`, as a security enhancement strategy in §3.2.6.

### 3.2.1 Design of `kth_jit`

Our `kth_jit` kernel thread is designed to perform periodically re-randomization operation of the L2L table. It is invoked at the load time of the binary, and then constantly wakes up to perform the re-randomization by shuffling the logic page numbers in our L2L table. We use the Fisher-Yates shuffle algorithm [28] for our purpose, and the random source is from the kernel function `get_random_bytes`.

In addition to shuffling the L2L table, `kth_jit` also needs to alter the page table entries (PTEs) according to the current mapping. For example, if logical page $L1$ in the left column of L2L table corresponds to $L2$, the kernel thread will replace $L2$'s PTE with the PTE of $L1$. After the kernel thread repeats this operation for every logical page in the auxiliary table, the page table randomization is completed. We then use kernel function `flush_tlb_range` to flush the part of TLB accordingly.

To simplify the page table handling, we currently keep all the code pages into memory at program load time (using `mlock`) to avoid the complexities that will be otherwise introduced by swapping.

### 3.2.2 Design of `sys_jit` & CHAMELEON stub

As illustrated in our working example, our `sys_jit` needs to determine the original target address to get the original page number, look up the L2L table to get the new page number, and finally redirect the control flow to the destination address.

**Determining the Target Address.** To decide the after-randomization address, we need to know its original before randomization address such that we can query our L2L table. For direct `call/jmp`, their target addresses are explicitly given in the binary code, and we can directly fetch them from the *target address* argument of `sys_jit`. For those dynamically computed target addresses (e.g., load from a memory or register) such as `call/jmp *ecx`, we will pass their run-time address to our `sys_jit` argument. In most cases, the run-time addresses are before-randomization addresses determined by the compiler or loader, for example, jump table, virtual function table. However, there are some special cases that run-time addresses are after-randomization addresses (calculated from `eip`). If `sys_jit` receives such address and looks up L2L table, the result will be incorrect. To solve this problem, we restore the `eip` to before-randomization address when it is saved on user stack (details is in §3.2.5).

There are also some other target addresses we have to decide at runtime. Specifically, (1) if the inter-page instruction is a conditional jump, we need to determine the target address based on `eflags` and/or `ecx` (note instruction such as `jcxz/jecxz` will jump if `ecx` is 0). (2) If the instrumented instruction is `ret`, we need to get the before-randomization return address from the kernel (§3.2.6).

**Querying L2L Table.** `sys_jit` will query L2L table for two tasks: (1) given the before-randomization address to get the after-randomization target address, and (2) given the after-randomization address to get the before-randomization return address. For task (1), since the left column of the L2L table represents page number, it just needs to convert logical address to page number (by `addr & PAGE_MASK`) before looking up the L2L table, use this page number as the entry to find the after-randomization page number. For task (2), similar to task (1), the system call inquires the L2L table to get the before-randomization page number and calculate the before-randomization return address. Note that after-randomization target address and before-randomization return address will be saved at the reserved space by RSI unit (Recall §3.1.1).

**Redirecting the Control Flow – CHAMELEON Stub.** Before redirecting the control flow to a target address, our `sys_jit` has to perform some restoration. Considering the fact that this restoration procedure is exactly the same for all types of instrumentations, we simply include the restoration procedure inside a wrapper function and we call it CHAMELEON stub. Therefore, right before the `sys_jit` finishes its execution, it will modify the saved return address (which is in the top of user stack) with the entry address of CHAMELEON stub. Then the control flow will directly go to the CHAMELEON stub, which will `popa` and `popf` to restore the registers. The CHAMELEON stub is also just-in-time randomized (resistant to just-in-time code reuse attacks) and we store them in a separate page.

After the context restoration, it is safe to change the control flow to the after-randomization target address. Here similar to ROP, we use a trick to set the target address. Specifically, instead of inserting a jump instruction after returning from the CHAMELEON stub, we insert a `ret` instruction in the Chameleon stub, and replace its return address with the after-randomization target address. Such replacement operation is achieved inside the `sys_jit` system call.

As shown in Figure 2, `sys_jit` modifies the return address in the current stack frame to be the entry address of the stub. In addition, `sys_jit` writes the after-randomization target address `0x8105f10` to a reserved space in the stack (❼). When `sys_jit` returns, the execution will be redirected to the stub (❽). Finally the stub returns to the after-randomization target address (i.e, `0x8105f10`) stored in the reserved stack space (❾).

### 3.2.3 Synchronization between `kth_jit` & `sys_jit`

Since both `kth_jit` and `sys_jit` need to access the shared L2L table, a proper synchronization between them is needed. On one hand, our `sys_jit` only needs to look up the shared table (i.e., read-only). On the other hand, our `kth_jit` randomizes and modifies the shared tabled (i.e., read-write). We leverage the kernel `rw_semaphore` to synchronize the accesses to L2L table. The system call will call `down_read()` and `up_read()` to hold and release the read lock and the kernel thread will call `down_write()` and `up_write()` correspondingly. Besides, we employ `schedule_timeout()` to implement the timed sleep of the kernel thread.

We have to set/reset the current executing page as no-re-randomize page in a white list, as such a different synchronization policy is needed. This is because the kernel thread will only read the information and the system calls will both read and write the information. As `sys_jit` and `kth_jit` have already been mutually excluded by the `rw_semaphore`, we only need to prevent multiple system calls (in a multithreaded program) from concurrently modifying the white list. Since the set/reset operations are supposed to be very quick, we choose the `spinlock`. It is worth noting that for a single-threaded process, the `spinlock` can actually be eliminated, which can potentially improve the performance.

Regarding multithreaded programs, they are naturally supported by our design, because threads in the same process share the same set of page table and they can concurrently issue `sys_jit` system calls which have already been appropriately synchronized with the kernel thread. In order to support multiple processes running at the same time, each process needs to be randomized timely and correctly. Thus, we associate a different kernel thread for every process and we dynamically allocate (i.e., `kmalloc`) the L2L table for each process. We store a reference to the L2L table in each process's `task_struct` so that the system calls can quickly identify the corresponding table to look up.

### 3.2.4 Position Independent Code

CHAMELEON does support securing the position independent code (PIC). In particular, during the static instrumentation phase of PIC, the flag argument of `sys_jit` is set to indicate the invoking code is PIC and which library it belongs to. In the kernel space unit, we take following actions. First, we hook the `mmap_pgoff` in the kernel space to get information of each loaded and instrumented dynamic library, including (1) the current process id; (2) the base address of the library; (3) the name of the library. Knowing the start offset and end offset of the code segment, `kth_jit` can understand the start address and end address of the library. Then the code pages of the library are put in the L2L table together with the code pages of the application. At the end of randomization cycle, `kth_jit` updates the GOT table for the library based on the newest L2L table. After that, when a `sys_jit` is invoked in PIC, the `sys_jit` checks the target address passed as an argument. If the target address is less than the base address of its containing library, we will add it with the base address.

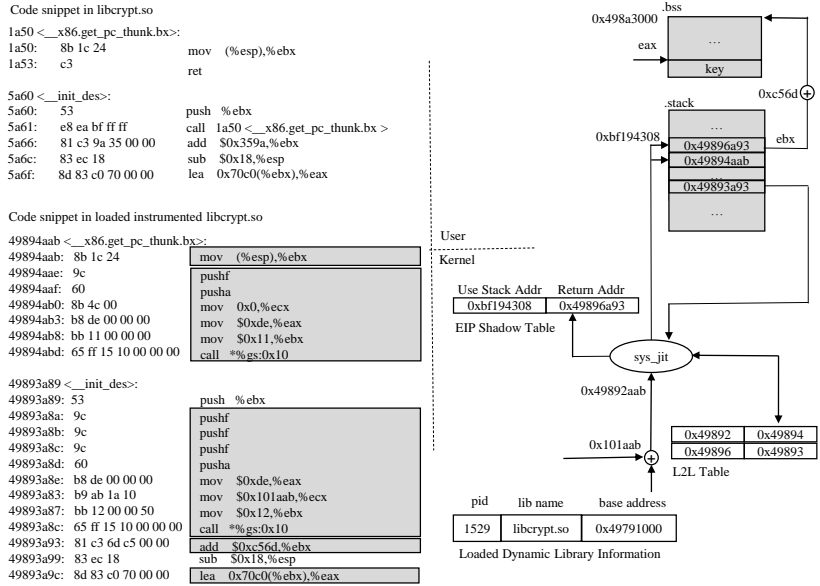

Figure 3: Dynamic Library libcrypt.so

We take the instruction `call 0x1a50` from `libcrypt.so` in Figure 3 as an example to illustrate how we handle PIC. After static instrumentation, `call 0x1a50` is replaced by a RGI unit (`0x49893a8a-0x49893a92`). The flag arguments `0x12` of `sys_jit` indicates the code is `call` instruction in `libcrypt.so`. After `libcrypt.so` is loaded into a process (e.g., httpd), the kernel will record the loaded dynamic library information: the process id (`0x1529`), library name `libcrypt.so` and base address `0x49791000`. At runtime, when `sys_jit` is invoked at address `0x49893a8c` with target address argument `0x101aab`, `sys_jit` will look up loaded dynamic library information, and add the base address `0x49791000` to get the target address `0x49892aab`. As we illustrated in §3.2.2, `sys_jit` will look up the L2L table, write the after-randomization target address `0x49894aab` in the reserved space in user stack, and transfer the control flow to `0x49894aab` (i.e., `__x86.get_pc_thunk.bx`).

### 3.2.5 Dynamically Computed Address

As discussed in §3.2.2, dynamically computed addresses could be before-randomization addresses or after-randomization addresses. Before-randomization addresses can be directly used to do L2L table query. So here we focus on the second type of dynamically computed addresses.

The reason for a dynamically computed address to be after randomization is that the saved `eip` on the user stack is used in the address computing. On a 32-bit Linux, `eip` cannot be directly accessed by any instruction. In two ways, `eip` will be indirectly accessed and saved to the stack: (1) `call` instruction saves the current `eip` on the stack as return value; (2) during signal handling, the kernel saves the `eip` in a signal frame (`sigframe`) by `setup_frame`, and later stores the user content by `sig_return`. We denote the addresses that dynamically computed from the saved `eip` on the stack as *EIP relative addresses*.

Taking the global variable (`key`) in `libcrypt.so` in Figure 3 for example, `key` is in the `bss` segment with a fixed offset `0x70c0`. To get the address of `key`: (1) the PIC code uses "`call 1a50`" to save the return address (i.e., address of "`add $0x359a,%ebx`") on the stack; then "`mov (%esp),%ebx`" assigns

Table 1: Statistics of the Instrumented Programs.

| Programs | Orig_Sizes (KB) | RSB_Sizes (KB) | Iter_Count | Orig_Page | RSB_Page | Padding (Bytes) | #total CFT in Orig | #RSI units |
|---|---|---|---|---|---|---|---|---|
| cat | 292 | 318 | 24 | 8 | 11 | 401 | 1652 | 194 |
| tail | 368 | 455 | 25 | 11 | 15 | 462 | 2298 | 240 |
| ls | 644 | 969 | 57 | 18 | 30 | 728 | 3901 | 428 |
| du | 664 | 979 | 56 | 20 | 31 | 892 | 4313 | 432 |
| cp | 684 | 986 | 56 | 19 | 32 | 1033 | 4078 | 452 |
| sort | 624 | 716 | 44 | 17 | 26 | 837 | 3455 | 403 |
| head | 228 | 204 | 21 | 6 | 8 | 370 | 1193 | 152 |
| base64 | 200 | 164 | 13 | 5 | 11 | 287 | 953 | 136 |
| expand | 176 | 142 | 17 | 5 | 9 | 123 | 771 | 118 |
| unexpand | 176 | 140 | 14 | 5 | 9 | 110 | 806 | 120 |
| paste | 176 | 130 | 16 | 5 | 9 | 263 | 780 | 120 |
| xz-5.1.4beta | 385 | 483 | 40 | 11 | 18 | 729 | 2325 | 216 |
| bzip2-1.0.6 | 402 | 643 | 39 | 16 | 22 | 396 | 2733 | 232 |
| ghttpd-1.4-4 | 92 | 65 | 6 | 3 | 5 | 89 | 444 | 56 |
| httpd-2.2.27 | 2867 | 3621 | 136 | 77 | 116 | 3072 | 17184 | 1945 |
| nginx-1.6.0 | 3788 | 4154 | 157 | 98 | 147 | 4017 | 22040 | 2765 |
| lighttpd-1.4.35 | 1024 | 1419 | 73 | 27 | 42 | 2053 | 5906 | 707 |

the return address to `ebx`, and "`add $0x359a,%ebx`" calculates the base address of `.bss` segment (the offset of "`add $0x359a,%ebx`" to `bss` segment is `0x359a`). Finally, "`lea 0x70c0(%ebx),%eax`" assigns the address of `key` to `eax`. The address of `key` is an EIP relative address. In the loaded instrumented `libcrypt.so`, the return address is `0x49893a93` which is after-randomization address. After adding the offset (`0xc56d`), the base address of `bss` segment will be calculated as `0x498a0000`, which is not the correct base address (`0x498a3000`). Note if we use the before-randomization return address `0x49896a93` to calculate the base address, we will get the correct result.

To handle EIP relative address, when saving `eip` on user stack, `sys_jit` or `setup_frame` (hooked by us) will save the before-randomization address. Take the working example in Figure 2 for example, `sys_jit` splits the return address (`0x8103063`) with a page number `0x8103` and offset `0x63` (❺), and then performs an inverse query of L2L table to get the page number of the before randomization return address, which is `0x8109` in this case (❻). Consequently, the before randomization return address will be `0x8109063`, and we should put this address in the reserved stack space. Similarly, we save before-randomization return address `0x49896a93` on the stack in Figure 3.

### 3.2.6 Security Enhancement

**Secure the saved `eip`.** To handle EIP relative addresses, we save the before-randomization value of `eip` on the stack. The attack (e.g., ROP attack [40] and SROP [10]) might attempt to modify these saved values to the addresses of malicious gadgets (addresses of these gadgets can be analyzed offline). To secure the saved `eip`, we rely on the kernel space to provide protection. In the kernel space, CHAMELEON maintains an EIP shadow table in the kernel space, which contains the before-randomization value of saved `eip` and the saving location. For instance, in Figure 3, the return address (`0x49896a93`) and its saving location `0xbf194308` are stored in the EIP shadow table. When `sys_jit` (replacing the `ret`) or `sig_return` (hooked by us) restores the `eip`, it (1) looks up the EIP shadow table to get the `eip` ; (2) gets the after-randomization address through L2L table; (3) restores the `eip` to the after-randomization address. In this way, no matter how the saved `eip` on stack is crafted, the control flow is valid because it is based on `eip` saved in the EIP shadow table in the kernel space.

**How can we disable reuse of `sys_jit`.** An astute attacker might be interested in extending "one-two punch" ROP attacks with reusing `sys_jit`. However, our approach makes reuse of `sys_jit` nearly impossible. First, it is difficult for an attacker to constantly identify the locations of RSI units since pages containing RSI units are actively and periodically randomized. Second, even if an attacker can always get the correct locations of RSI units, the attacker needs to fake the arguments of `sys_jit`, especially the target address argument. There are three types of target address: (1) hard-coded label; (2) return address; (3) register or register involved memories (e.g., `jmp %eax` ). With DEP enabled, it is hard for the attacker to craft the hard-coded label in code segment. So it is difficult to reuse `sys_jit` with the first type of return address argument; in our design, all return addresses are maintained in the kernel, which is impossible to be modified by an attacker without root privilege. To this end, it is impossible for the attacker to reuse `sys_jit` with the second type of return address argument; to reuse `sys_jit` with the third type of return address argument, the attacker needs to modify the registers to make it either point to the gadget address or the memory that contains the gadget address. However, register modification requires execution of specific gadgets (e.g., dispatch gadgets) [12, 9]. To execute any gadgets without reusing `sys_jit` has been explained to be impossible when our defense principle is enforced by VAPTR. In summary, it is never easy for an attacker to reuse `sys_jit`.

## 4 Evaluation

We have implemented CHAMELEON. Our user level static binary instrumentation component is implemented on the top of DyninstAPI library (8.1.2) [25], with additional 9000 lines of C++ code, and our kernel component (VAPTR) and the `sys_jit` are implemented atop Linux kernel 3.16.0 with over 1100 lines of C code added. In this section, we present our evaluation results.

We first evaluated the effectiveness of CHAMELEON with a number of JIT-ROP attacks (§4.1), and then tested the performance of programs protected by CHAMELEON (§4.2). The experiment suit consists of 113 programs: among them, 107 programs are from the `coreutils` toolchain (8.21) [16], 2 programs from commonly used compression software (`xz`, `bzip2`), and 4 programs from web servers (`ghttpd`, `httpd`, `nginx`,`lighttpd`). All the tests are conducted on an Intel(R) Core2 Duo CPU E8400, 3.00GHz desktop with 2GB physical RAM.

### 4.1 Effectiveness
#### 4.1.1 Static Binary Instrumentation

We successfully instrumented all of the 113 binaries. Due to space limitation, Table 1 only presents the statistics for 17 programs, and among them 11 are from the coreutils (the criteria of selection is based on the run time, and these programs take longer time that all others). The first column is the names of the program; the $2^{nd}$ column and $3^{rd}$ column are the size of original binary and the instrumented binary, respectively. We notice that the average increase of the binary file size is 47.2%; the $4^{th}$ column reports the number of iterations our instrumentation takes. In general the iteration count depends on the size of the binary. In particular, the largest iteration count is 157 (`nginx`) and the smallest one is 6 (`ghttpd`); the $5^{th}$ column reports the number of *code* pages in the original binary and the $6^{th}$ column reports that of the instrumented binary (i.e., the RSB). About 50% more code pages are introduced; the $7^{th}$ column reports the size of padding inserted in the instrumented binary; the $9^{th}$ column reports the number of RSI units inserted to the instrumented binary. For all of the test programs, we inserted on average 31 padding bytes and 21 RSI units per page.

To show how many control flow transfer (CFT) instructions in the original binary are instrumented by RSI units, we compared the total number of CFT instructions in the original binary (the $8^{th}$ column) with the number of RSI units (the $9^{th}$ column), and found that 24.7% CFT instructions in the original binary are instrumented. To be short, we will use iCFT (inter-page control flow transfer) instructions to represent inter-page instructions.

Table 2: Security Evaluation of CHAMELEON

| Programs | wo/ CHAMELEON on original binary | | | | | w/ CHAMELEON on instrumented binary | | |
|---|---|---|---|---|---|---|---|---|
| | Memory Disclosure | | Gadgets Discovery And Just-in-time Compilation | | Gadgets Execution | Memory Disclosure | Gadgets Discovery And Just-in-time Compilation | Gadgets Execution |
| | # of Pages | Time (ms) | # of Gadgets | Time (ms) | # of Gadgets | # of Pages | # of Gadgets | # of Gadgets |
| cat | 5 | 255.69 | 710 | 2852.25 | 15 | 1 | 89 | 0 |
| tail | 7 | 314.95 | 1291 | 3589.72 | 22 | 1 | 116 | 0 |
| ls | 10 | 319.05 | 2137 | 3669.57 | 22 | 1 | 154 | 0 |
| du | 10 | 320.35 | 2083 | 3601.66 | 22 | 1 | 122 | 0 |
| cp | 14 | 325.83 | 2251 | 3755.16 | 21 | 1 | 78 | 0 |
| sort | 12 | 324.63 | 2562 | 3698.74 | 22 | 1 | 94 | 0 |
| head | 4 | 314.29 | 699 | 3561.91 | 22 | 1 | 66 | 0 |
| base64 | 3 | 308.36 | 614 | 3442.44 | 22 | 1 | 65 | 0 |
| expand | 3 | 318.44 | 545 | 3552.72 | 22 | 1 | 79 | 0 |
| unexpand | 3 | 320.47 | 527 | 3552.07 | 22 | 1 | 78 | 0 |
| paste | 4 | 320.14 | 662 | 3578.12 | 22 | 1 | 65 | 0 |
| xz | 10 | 321.74 | 1793 | 3660.82 | 22 | 1 | 76 | 0 |
| bzip2 | 13 | 392.80 | 1788 | 3927.24 | 23 | 1 | 142 | 0 |
| ghttpd | 1 | 8.56 | 96 | 0 | 0 | 1 | 33 | 0 |
| httpd | 27 | 314.54 | 3978 | 3182.25 | 21 | 1 | 177 | 0 |
| nginx | 86 | 447.53 | 13393 | 4675.23 | 24 | 1 | 184 | 0 |
| lighttpd | 32 | 345.78 | 5452 | 3629.13 | 24 | 1 | 104 | 0 |

### 4.1.2 Defense against the JIT Code Reuse Attack

**Attack Construction.** To test the effectiveness of CHAMELEON against JIT code reuse attacks, we first construct a set of JIT code exploits against all the original binaries as well as the instrumented binaries. Since a tested program may not have the desired vulnerability, similar to how JIT-ROP [41] evaluates their techniques, we create a shared library that is linked by each tested program to simulate the JIT-ROP attack. This library contains our exploit routine that will perform memory disclosure and gadget collection, just-in-time gadgets compilation, and gadgets execution. To hijack the program control flow to execute the routines in our shared library, we directly patch the entry address of the program and let it directly execute our attack payload.

**Attack Results.** We run the constructed exploit against both the original binaries and the instrumented binaries. Table 2 presents our experimental result. From the $2^{nd}$ column to the $6^{th}$ column, we present the statistics without CHAMELEON, and from the $7^{th}$ column to the $9^{th}$ column, we present the statistics with CHAMELEON.

Regarding the original binary programs without CHAMELEON protection, all of them can be successfully attacked except the `ghttpd`, which has too few (only 96) gadgets to perform even a simple attack (e.g., open a shell). Nevertheless, we can still complete the memory disclosure task against it. Since the time of "gadgets execution" is very short (no more than $1\mu s$), we only show the time consumed by the other two tasks in the table. On average, 14 pages are scanned (between 1 and 86), which take 310.18ms. On average, 2387 gadgets are discovered (between 96 and 13393) and it takes 3620.56ms to discover and compile these gadgets. 15-24 selected gadgets are executed to complete the attack.

When attacking instrumented binaries, we run with three different re-randomization time intervals: 1ms, 10ms, 100ms, and all of the exploit routines failed without any exception. The reasons for choosing such re-randomization intervals are discussed in §5. We observe that all failures share some common characteristics. First, only one page can be scanned no matter which binary is being attacked, and that page is the attack starting page. This is because we have eliminated all the iCFT instructions to prevent the attack from disclosing other pages. Second, the number of discovered gadgets is 99 on average, and all of them are on the same page (i.e., the attack starting page). The number of discovered gadgets is sharply reduced by 95.9% compared to the original binary. Third, for all the test programs, the gadget compilation cannot succeed because the exploit routine failed to identify sufficient amount of gadgets in order to achieve a simple task (i.e., opening a shell). Finally, to evaluate the effectiveness of our just-in-time randomization, we compile a very limited number of discovered gadgets (2-10 gadgets) and executed them to do some specific operations (e.g., data movement). The execution of the selected gadgets cannot be completed and the exploit routines always crashed. We find that the crash of the gadgets execution is due to our re-randomization.

Table 3: Gadget Statistics

| Programs | Original Binary #Gadgets | Instrumented Binary # Gadgets |
|---|---|---|
| cat | 1200 | 401 |
| tail | 1979 | 984 |
| ls | 4029 | 2001 |
| du | 3963 | 1793 |
| cp | 3613 | 918 |
| sort | 3645 | 1377 |
| head | 1057 | 356 |
| base64 | 1006 | 341 |
| expand | 906 | 428 |
| unexpand | 878 | 259 |
| paste | 815 | 231 |
| xz | 1964 | 859 |
| bzip2 | 2210 | 1225 |
| ghttpd | 281 | 101 |
| httpd | 12893 | 6305 |
| nginx | 16161 | 7272 |
| lighttpd | 4967 | 2185 |

Note that if the attacker uses side channel-based techniques [7, 38] to collect the gadgets on-the-fly, it generally will need more time. For example, the fastest exploit of BROP [7] takes 1 minute against nginx. To exploit the Apache Web Server, [38] takes 30.58 sec under traditional ASLR [43], and even 1 week under fine-grained ASLR protection. During such a period, CHAMELEON would perform re-randomization for hundreds of times. If the re-randomization interval is 100ms, Chameleon will perform more than 300 times of re-randomization during 30.58 sec. As such, we believe CHAMELEON has the capability to defeat these side channel attacks.

Table 4: Runtime Statistics of the Instrumented Programs

| Programs | Commands or Benchmarks | #RSI unit executions | #unique RSI units involved | $> 10^6$ | $> 10^5$ | $> 10^4$ | $> 10^3$ |
|---|---|---|---|---|---|---|---|
| cat | cat a 1.3MB file | 1,039 | 24 | 0 | 0 | 0 | 0 |
| tail | tail the last 1M lines from a 3.0MB file | 370 | 33 | 0 | 0 | 0 | 0 |
| ls | ls a directory that contains 100K files | 20,260 | 150 | 0 | 0 | 0 | 5 |
| du | du a 4GB directory trees | 4,716,284 | 610 | 0 | 3 | 64 | 220 |
| cp | copy 430MB files | 4,347,857 | 280 | 0 | 7 | 76 | 125 |
| sort | sort a 1.3MB file | 9,820,150 | 224 | 2 | 20 | 24 | 25 |
| head | head the first 1M lines from a 3.0MB file | 748 | 18 | 0 | 0 | 0 | 0 |
| base64 | base64 a 3.0MB file | 1,951 | 14 | 0 | 0 | 0 | 0 |
| expand | expand a 3.0MB file | 119,009 | 13 | 1 | 1 | 1 | 1 |
| paste | paste a 3.0MB file | 241,688 | 23 | 0 | 0 | 4 | 4 |
| unexpand | unexpand a 3.0MB file | 119,811 | 17 | 0 | 0 | 2 | 2 |
| xz-5.1.4beta | compress a 3.0MB file | 1,757 | 110 | 0 | 0 | 0 | 0 |
| bzip2-1.0.6 | compress a 12.3MB file | 51,576,587 | 365 | 14 | 26 | 190 | 206 |
| ghttpd-1.4-4 | wget http://ip:port/index.html | 213 | 15 | 0 | 0 | 0 | 0 |
| httpd-2.2.27 | ab -k -n50000 -c100 http://ip:80/index.html | 1,584,667 | 427 | 0 | 1 | 21 | 245 |
| nginx-1.6.0 | ab -k -n50000 -c100 http://ip:80/index.html | 4,240,472 | 1,066 | 0 | 12 | 285 | 294 |
| lighttpd-1.4.35 | ab -k -n50000 -c100 http://ip:80/index.html | 2,339,981 | 750 | 0 | 12 | 76 | 311 |

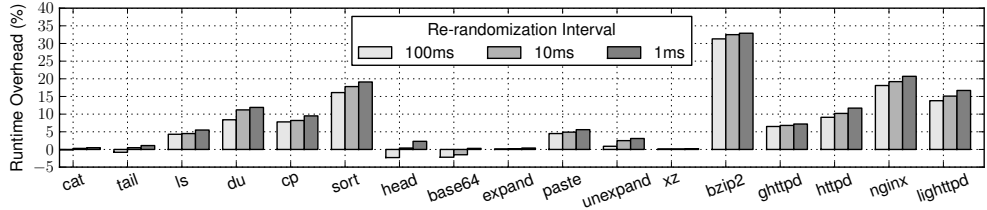

Figure 4: Runtime Overhead

**Worst Case Analysis.** In our simulated attack, we assume the attacker follows the steps in JIT-ROP [41]. Next, we elevate the attacker's potential to an extreme case, namely, every gadget could be used by the attacker. However, by comparing the original binary and the instrumented binary in terms of the total number of gadgets, we can find that our RSB has significantly less gadgets as shown in Table 3. More specifically, compared with the average 3621 gadgets in the original binaries, the instrumented binaries only have 1590 gadgets, with a significant decrease of 56.1%. It is worth mentioning that the remaining gadgets are all unintended gadgets, and all the intended gadgets have been removed by the nature of our technique (recall that all the `ret` instruction in the original binary has been replaced).

## 4.2 Runtime Overhead

Next, we measured the runtime overhead for the tested programs. For each program, we measured the followings: (1) runtime statistics of the instrumented binaries without no-security-loss optimization, (2) runtime overhead with no-security-loss optimization, and (3) the effectiveness of optimization. For each test, we ran the program 1,000 times and calculated the average.

Table 5: Evaluation Result of Our Performance Optimization

| Program | Before Optimization | | | | | | After Optimization | | | | | |
|---|---|---|---|---|---|---|---|---|---|---|---|---|
| | Overhead | # of RSI unit executions | # of RSI unit executions | | | | Overhead | # of RSI unit executions | # of RSI unit executions | | | |
| | | | $>10^6$ | $>10^5$ | $>10^4$ | $>10^3$ | | | $>10^6$ | $>10^5$ | $>10^4$ | $>10^3$ |
| ls | 9.2% | 20,260 | 0 | 0 | 0 | 5 | 4.3% | 10,264 | 0 | 0 | 0 | 1 |
| du | 16.4% | 4,716,284 | 0 | 3 | 64 | 220 | 8.4% | 3,400,698 | 0 | 0 | 44 | 154 |
| cp | 10.4% | 4,347,857 | 0 | 7 | 76 | 125 | 7.8% | 3,271,124 | 0 | 2 | 62 | 97 |
| sort | 35.5% | 9,820,150 | 2 | 20 | 24 | 25 | 16.1% | 4,972,380 | 1 | 11 | 11 | 11 |
| expand | 3.1% | 119,009 | 1 | 1 | 1 | 1 | 0.1% | 12 | 0 | 0 | 0 | 0 |
| unexpand | 4.8% | 119,811 | 0 | 0 | 2 | 2 | 0.9% | 60,967 | 0 | 0 | 1 | 1 |
| paste | 10.5% | 241,688 | 0 | 0 | 4 | 4 | 4.5% | 120,864 | 0 | 0 | 2 | 2 |
| bzip2 | 136.7% | 51,576,587 | 14 | 26 | 190 | 206 | 31.3% | 11,713,541 | 1 | 12 | 174 | 189 |
| httpd | 14.1% | 1,584,667 | 0 | 1 | 21 | 245 | 9.1% | 884,289 | 0 | 0 | 3 | 181 |
| nginx | 26.4% | 4,240,472 | 0 | 12 | 285 | 294 | 18.1% | 2,389,676 | 0 | 8 | 203 | 212 |
| lighthttpd | 21.2% | 2,339,981 | 0 | 12 | 76 | 311 | 13.8% | 1,338,676 | 0 | 4 | 63 | 242 |

**Runtime RSI Execution Statistics.** Table 4 shows the runtime statistics of 17 selected programs when running without no-security-loss optimization. The $2^{nd}$ column shows the specific shell commands or benchmarks. The runtime overhead is mainly due to the RSI units, since the padding (`NOP` sequence) will introduce negligible overhead. The statistics include the total number of executed RSI units (the $3^{rd}$ column) and the number of unique RSI units involved (the $4^{th}$ column). As we can see, `ghttpd` executed the smallest number of RSI units (213), while `bzip2` executed the largest number of RSI units (51,576,587). Note that `bzip2` used an extremely large number of loops.

We denoted the RSI units which are executed more than 1,000 times as frequently executed RSI units. We further decomposed the frequently executed RSI units into four categories in Table 4. For example, ">$10^6$" means the number of unique RSI units that are executed over $10^6$ times per each. `bzip2` has 14 such RSI units, `expand` has 1 such RSI unit and `sort` has 2 such RSI units. Although the number of the frequently executed RSI units shares only 21.3% of all RSI units, the frequently executed RSI units consume 82.6% of the total number of executions. In fact, after applying the no-security-loss optimization presented in §3.1.3, we can safely remove some of the frequently executed RSI units.

**Runtime Overhead.** For the 113 binaries with no-security-loss optimization, we run four sets of test. We first run the original program. Then we run the instrumented program with three different re-randomization intervals, namely, $100ms$, $10ms$, and $1ms$. And for each set of test, we run the executable 1,000 times. The average runtime overhead under the three intervals are 11.1%, 12.1%, and 12.9%, respectively. When we calculate the average runtime overhead, only the binaries that have an execution time longer than a threshold (i.e., 500ms) are considered. We get 44 binaries in this category. Fig. 4 shows the runtime overhead measurement for the 17 selected programs.

By correlating with Table 4, we can see that the runtime overhead is heavily correlated with the number of executed RSI units and the number of frequently executed RSI unites. The higher the numbers are, the more runtime overhead will be introduced. In practice, we noticed that if the number of executed RSI units is less than 2,000, the performance overhead is negligible (e.g., 0.1% for `xz`). Some instrumented executables (e.g., `cat`, `tail`, `base64` and `head`) even get better performance than the original binaries. One possible reason is that improved memory locality is achieved due to our instrumentation.

Note that about 60% test programs in the coreutils run less than 500ms and execute only 10-20 RSI units. Although the absolute number of RSI units is lower, the relative time consumption of RSI units can be higher. In addition, the initialization overhead (including the creation of `kth_jit` kernel thread and the L2L Table) has a greater impact on short-lived programs. As a result, many of the small command-line programs, such as `hostid`, `logname`, and `whoami`, have runtime overhead of up to 30%.

In addition to the coreutils, we also test three web server programs: `httpd-2.2.27`, `nginx-1.6.0`, and `lighttpd-1.4.35`. We use the ApacheBench (ab) to simulate 50,000 HTTP GET requests and evaluate the performance when setting the re-randomization interval to $100ms$. Without optimization, we observe 1,584K, 4,240K, 2,339K executions of RSI units, respectively, for these programs; with optimization, we observe 884K, 2,389K, 1,338K executions of RSI units, yielding a reduction of 44.2%, 43.7% and 42.8%, respectively. Consequently, as shown in Figure 4 and Table 5, the runtime overhead is reduced from 11.7% to 9.1%, 20.7% to 18.1%, 16.7% to 13.8%, respectively, for these programs.

**Dynamic Library Evaluation.** We also include a number of dynamic libraries in our test suit: `libcrypt.so`, `libutils.so` and `libssl.so`. In our experiments, we successfully instrumented all these libraries. In particular, we select the instrumented `libssl.so` to evaluate how it affects the performance of the web server software (`httpd`, `nginx`, `lighttpd`). Note that `libssl.so` is widely used in web applications to provide https service and should be a representative candidate for our evaluation. With our optimization performed, CHAMELEON instruments $3,843$ inter-page instructions in `libssl.so`, which are 24.5% of total control flow instructions ($15,684$). For each instrumented web application, we link it with the original `libssl.so` and the instrumented `libssl.so` to evaluate the performance difference. By setting the HTTPS benchmark as `ab -k -n50000 -c100 https://ip:443/index.html` and setting re-randomization interval as 10*ms*, we observe that CHAMELEON introduces 10.8%, 23.1% and 16.3% performance overhead, respectively. We look further to see how many RGI units get executed, and find that 948K, 2,798K, and 1,433K RGI units executed in `httpd`, `nginx` and `lighttttpd`, respectively.

**Effectiveness of Our Optimization.** To measure the effectiveness of the optimizations presented in §3.1.3, we select 11 programs that contain frequently executed RSI units to compare the runtime overhead before and after the optimization, using $100ms$ re-randomization interval. Table 5 represents the results.

Overall, the average overhead is reduced by 59.9% (from 26.2% to 10.5%). It turns out that the program `bzip2` benefits the most from the optimization. The overhead is reduced from 136.7% to 31.3%. As the performance improvement primarily comes from the reduction in the number of executed RSI units, we collect the number of executed RSI units before and after optimization. Although the reduction of unique RSI units is 24.2%, the gross reduction rate of executed RSI units is 64.4%. This indicates that our optimization reduces the frequently executed RSI units to a large extent.

## 5 Discussions

**How long should each randomization cycle be.** The length of each randomization cycle is critical to our approach. The cycle being too long would leave a door for the attack to get in; the cycle being too short would introduce high overhead. It is hard to theoretically identify a secure length of each randomization cycle to stop all attacks. We determine the length of each randomization cycle via mimic JIT-ROP attacks against the program to be protected. Details of experiments are presented in §4.1 and results are presented in Table 2. The shortest time to conduct a JIT-ROP attack is over 3,000 *ms* (`cat`). Note that in real-world attacks, the time should be much longer, due to issues such as network delay and vulnerability exploitation. So we provide three options for the randomization cycle: 1 *ms*, 10 *ms*, and 100 *ms*. In fact, if necessary, we can set the randomization interval to be much shorter.

**Compatibility between Chameleon and ASLR.** ALSR is a commonly used defense technique built-in modern Linux kernels. We tested whether CHAMELEON is compatible with ASLR using the experiment suit. The results show that all programs protected by CHAMELEON run correctly when ASLR is enabled.

## 6 Related Work

Since the introduction of data execution prevention (DEP), attackers cannot execute the injected code any more, and they have to reuse the existing code for the malicious purpose. Over the past decades, we have witnessed this code reuse attack evolved from return-into-libc [32, 22], to ROP/JOP [40, 37, 12, 9, 11], to the most recent just-in-time ROP [41, 7, 38]. In this section, we discuss the defense against these attacks and compare CHAMELEON with the most related one.

**Defense based on gadget patterns.** Due to the nature of the diverting the control flow and executing the existing program code, these code reuse attacks often expose distinctive patterns from benign program execution. For instance, they often have short code snippets (i.e., usually less than five instructions); there is no corresponding "call" for the "ret" instructions; the control flow transfer heavily depends on the "ret" instructions.

In light of this, DROP [14] detects ROP attack based on the heavy use of "ret" for control flow transfer. ROPDefender [21] is based on the matched pair of call and return. kBouncer [35] and ROPecker [15] is based on the length of the gadget as well as the sequence of consecutive gadgets. However, two recent efforts [20, 23] have demonstrated these pattern based approach can be broken. Therefore, there are other efforts such as G-Free [33] and Return-less [29] proposing to remove the "ret" instructions so as to block the ROP attack.

**Defense based on randomization.** Since attackers have to know the address to which they can divert the control flow, another appealing defense is to randomize the locations of program code such that the address identified by attackers will fail. A large number of randomization techniques have been proposed, and they differ at the granularity such as segment offset level [43, 4], page level [3], function level [27], basic block level [34], and instruction level [45, 24]. Also, the randomization can occur at compile time (e.g., [6]) or load-time (e.g., [45, 4, 3]).

However, most of the existing ASLR only performs a one-time randomization. It thus leads to the JIT-ROP attacks [41, 7, 38]. A natural response would be to keep re-randomizing the program code. An early attempt is execution path randomization proposed by Isomeron [19]. Basically, with dynamic binary instrumentation, Isomeron keeps a copy of original code and the other copy of randomized code at runtime and randomly flips to execute either the original code or randomized code at function granularity. While CHAMELEON shares similar motivation of re-randomization, the substantial difference are (1) Isomeron does not follow our new defense principle, in other words, address-code associations are still kept under execution path randomization. It is possible that the attacker can hijack the control flow to execute JIT-ROP attacks without triggering the path execution randomization (function call). (2) Isomeron is dynamic binary instrumentation based, and it suffers from high performance overhead (2-5X); whereas CHAMELEON is much more practical and it uses static binary instrumentation with a number of optimization techniques, leading to around 10% performance overhead.

**Other Defenses.** With the same motivation of JIT code reuse attack, an approach named "Execute-no-Read" (XnR) is recently proposed [2]. The key idea is to ensure that code can still be executed by the processor, but cannot be read as data, which essentially forfeits the self-disassembly that is necessary for JIT code reuse attacks. However, gadget collection can still be done without "reading code as data". For example, the recently proposed BROP attack [7] can utilize a side channel (i.e., program crashes or not) to probe a valid gadget, without the need to actually read and disassemble the code. More advanced side channel attacks that do not require crashing the application have also been proposed [38]. It has been shown that a couple of fault analysis and timing channels can allow a remote attacker to learn how code has been diversified without needing to read the code. In both cases, XnR is incompetent to defeat the code reuse attacks. In comparison, CHAMELEON does periodical re-randomization. No matter how a gadget is collected, its logical address could have already been changed at the time of gadget execution. In this sense, CHAMELEON provides a more fundamental mechanism to defend against JIT code reuse attacks.

# 7 Conclusion

We have presented a new principle to instantly obsolete the address and code mapping to defeat the "one-two punch" code reuse attack. We have instantiated this principle with a novel technique called *virtual address page table remapping* and implemented in a system CHAMELEON, to periodically re-randomize the locations of code pages on-the-fly. To achieve this, we have developed a set of techniques including iterative instrumentation technique that instruments a to-be-protected binary program to generate a re-randomization suitable binary, runtime virtual page shuffling, and function reordering and instruction rearranging optimizations. The experiment results with over a hundred binary programs show that all of "one-two punch" code reuse attacks are defeated by our system. The re-randomization interval can be set as short as 100*ms*, 10*ms* or 1*ms*, and the experiment results show that CHAMELEON only introduces on average 11.1%, 12.1% and 12.9% performance overhead under these settings.